\documentclass[prb,twocolumn,superscriptaddress,showpacs,amsmath,amssymb]{revtex4-2} 
\usepackage{graphicx}
\usepackage{amsmath}
\usepackage{dcolumn}
\usepackage{bm}
\usepackage{hyperref}

\begin{document}

\title{ All-optical generation of Abrikosov vortices by the Inverse Faraday Effect }

\author{V.~D.~Plastovets}
\affiliation{Universite de Bordeaux, LOMA UMR-CNRS 5798, F-33405 Talence, France}
\affiliation{Institute for Physics of Microstructures, Russian
Academy of Sciences, 607680 Nizhny Novgorod, GSP-105, Russia}
\author{I.~D.~Tokman}
\affiliation{Institute for Physics of Microstructures, Russian
Academy of Sciences, 607680 Nizhny Novgorod, GSP-105, Russia}
\author{B.~Lounis}
\affiliation{Universite de Bordeaux, LP2N, F-33405 Talence, France}
\affiliation{Institut d'Optique \& CNRS, LP2N, F-33405 Talence, France}
\author{A.~S.~Mel'nikov}
\affiliation{Institute for Physics of Microstructures, Russian
Academy of Sciences, 607680 Nizhny Novgorod, GSP-105, Russia}
\affiliation{Lobachevsky State University of Nizhny Novgorod,
603950 Nizhny Novgorod, Russia}
\author{A.~I.~Buzdin}
\affiliation{Universite de Bordeaux, LOMA UMR-CNRS 5798, F-33405 Talence, France}

\date{\today}

\begin{abstract}

Within the framework of the time-dependent Ginzburg-Landau theory we show that circularly polarized THz or far-infrared radiation induces a dc supercurrent that influences the dynamics of vortex-antivortex pair formation in a mesoscopic superconductor undergoing rapid thermal quenching. The Lorentz force arising from the supercurrents promotes vortex-antivortex separation and allows survival of the vortex polarity defined by the helicity of light. 
Based on this idea, we propose a two-stage irradiation scheme that provides a powerful method for controlled all-optical generation of Abrikosov vortices.

\end{abstract}

\maketitle

\section{Introduction}
Optical manipulation of Abrikosov vortices is an attractive topic due to the simplicity of its experimental implementation and its perspectives for superconducting optoelectronic devices.
A simple mechanism of the light - vortex interaction originates from the local heating of superconductor by a tightly focused laser beam \cite{SEMENOV2001349, doi:10.1063/1.3374636, PhysRevB.85.024509, HotSpot}. 
The induced thermal gradient $\nabla T$ \cite{Sergeev_2010,https://doi.org/10.48550/arxiv.2203.12433} offers a possibility of fast and precise manipulation of individual Abrikosov vortices, demonstrated recently in Ref. \cite{Veshchunov2016}. 
Interestingly, a strong laser pulse by itself is able to generate vortices in superconductors by the Kibble-Zurek mechanism. 
First introduced in cosmology \cite{1975ZhETF673Z, Kibble_1976} and later generalized to the systems with broken U(1) symmetry \cite{Zurek1985, PhysRevLett.83.116, ARANSON2000129}, this mechanism can describe the formation of topological defects during rapid thermal quench after heating the sample with a thermal pulse.
Such a scenario has been observed in a superfluid helium \cite{Bauerle1996, Ruutu1996, VOLOVIK2000122}, superconducting systems \cite{PhysRevB.80.180501, PhysRevLett.91.197001,PhysRevLett.104.247002,Golubchik2011,PhysRevB.63.184520} and cold atoms condensate \cite{Weiler2008,PhysRevLett.128.150401}.
In superconductors, Kibble-Zurek mechanism always results in the creation of vortex-antivortex pairs which should annihilate during the post-quench dynamics.
The presence of pinning centers in the superconductor prevents the annihilation process making it possible to experimentally visualize the generated vortex-antivortex pairs \cite{PhysRevLett.104.247002, Golubchik2011}. 
In order to generate the vortex with a desired polarity at a desired position, one can use a focused laser pulse, which initiates the local rapid quench of the superconductor in the presence of a weak magnetic field (see Ref. \cite{doi:10.1021/acs.nanolett.0c02166}). 
The combined effect of the thermal force $f_T\sim-\nabla T$ and the Lorentz force arising from Meissner currents $f_L\sim j_M$ is able to separate vortex-antivortex pairs formed after quench with further transfer of desired polarity at the position of the laser spot and expelling the opposite fluxes to the edges of the superconductor.

The aim of this study is to consider an optical mechanism of inducing supercurrent, which can be used for a separation of vortex-antivortex pairs in a superconductor.
The basic idea is to replace an external magnetic field with a light radiation carrying a nonzero angular momentum. 
For instance, transfer of the \textit{orbital} angular momentum from Laguerre-Gaussian mode to the trapped Bose-Einstein condensate can excite persistent currents, which have been observed experimentally \cite{PhysRevLett.110.025301, PhysRevLett.99.260401}.
On the other hand, it is expected that the electromagnetic wave with a \textit{spin} angular momentum or just characterized by the circular polarization of a given helicity $\sigma_{\pm}$ can excite the circulating dc currents in a superconductor.
This problem is very similar to the so-called Inverse Faraday Effect (IFE)  consisting in generation of the magnetic moment in the sample irradiated by a circularly polarized electromagnetic wave \cite{RevModPhys.82.2731, Kirilyuk_2013}. 
In the case of a superconducting system, the light-induced dynamics of the order parameter comprises a nondissipative oscillatory contribution, which arises from the imaginary part of the order parameter relaxation time and creates the currents maintaining a nonzero magnetic moment \cite{PhysRevLett.126.137002}.  
It was theoretically shown \cite{PhysRevB.105.L020504, https://doi.org/10.1002/qute.202200054} that the interplay of the Kibble-Zurek mechanism and IFE in the case of a superconducting ring leads to the generation of the circulating current states with the rotation directions controlled by the external light polarization.

In the present paper we discuss the properties of IFE for the superconductor beyond the perturbation theory considered in Ref. \cite{PhysRevLett.126.137002} and demonstrate the possibility of using this effect for the vortex generation using the numerical solution of the time-dependent Ginzburg-Landau (TDGL) equations. 
More precisely we consider a mesoscopic superconductor homogenously quenched by a strong laser pulse and exposed to the circularly polarized THz radiation which provokes the IFE. 
We show that currents induced by the IFE lock up the vortices with the polarity dependent on the circular polarization helicity, thereby realizing the all-optical vortex generation. 

Note that the question of the vortex generation due to the direct transfer of the angular momentum to the superconducting condensate has already been addressed in Ref. \cite{Yokoyama_JPSJ}.
However, the analysis in Ref. \cite{Yokoyama_JPSJ} was based on the linearized TDGL equation which can not properly describe an essentially nonlinear problem of the vortex generation. The linearized model simply does not allow selecting the stable solutions.
Moreover, in Ref. \cite{Yokoyama_JPSJ} the purely real relaxation constant is considered, but in this case, the IFE is merely absent.
In contrast, the present study proposes different mechanism of the vortex generation, where the IFE plays the key role and the nonlinear TDGL model was used for the correct description of all stages of nucleation and evolution of vortices.


\section{Model}
The temporal evolution of the complex-valued order parameter $\psi({\bf r}, t)$ and the electric scalar potential $\varphi({\bf r}, t)$ in a superconductor square film is described by the modified TDGL equations:
\begin{gather} \label{GL1}
\tau_{\psi}\big(1+i\eta\big)\tilde{\partial}_t\psi=
\big\{a(t)-|\psi|^2 
-\xi^2{\bf D}^2\big\}\psi+f({\bf r}, t),  \\ \label{GL2}
\nabla^2 \varphi+\frac{\hbar}{2e\tau_{\text{GL}}}
\text{div}{\bf j}_s=0,   
\end{gather}
which are supplemented by the boundary conditions:
\begin{gather}
{\bf D}\cdot {\bf n}\Big|_S\psi=0, \quad 
\nabla\varphi\cdot {\bf n}\Big|_S=0.
\end{gather}
Here covariant operators ${\bf D}=(-i\bm{\nabla}-\frac{2\pi}{\Phi_0}{\bf A})$ and $\tilde{\partial}_t=( \frac{\partial}{\partial t}+\frac{2e}{\hbar}i\varphi)$ are introduced;  $\psi$ is expressed in terms of the equilibrium value of the order parameter in the absence of fields $\psi_0$; ${\bf A}$ is a vector potential; ${\bf j}_s=\text{Im}[\psi(\bm{\nabla}+i\frac{2\pi}{\Phi_0}{\bf A})\psi^*]$ is a supercurrent density; $a(t)=(T_c-T(t))/(T_c-T_0)$ is a temperature profile created by the homogeneous laser heating. 
The parameter $\tau_{\psi}=(\pi\hbar/8k_BT_c)/(1-T_0/T_c)$ is an order parameter relaxation time at the temperature $T_0$.
The crucial assumption which allows one to describe the dynamics of $\psi$ in terms of the equations (\ref{GL1},\ref{GL2}) is valid for gapless superconducting systems. Here we also introduce an imaginary part of the relaxation time of the order parameter proportional to a certain parameter $\eta$. 
This is the key parameter responsible for the IFE \cite{PhysRevLett.126.137002}, which appears due to the broken electron-hole symmetry \cite{kopnin2009theory, Larkin2008, PhysRevB.46.8376,Kopnin_sign}.
As a length unit we use here the coherence length $\xi=\xi_0/\sqrt{1-T_0/T_c}$ and the time unit is $\tau_{\text{GL}}= \tau_{\psi}/u$, where $u$ is the dimensionless characteristic time scale of the TDGL theory \cite{kopnin2009theory,Ivlev:1984}. 
Thermal fluctuations in a superconductor can be simulated using a delta-correlated stochastic force $f({\bf r}, t)$ \cite{Larkin2008,PhysRevB.59.9514,PhysRevB.71.214524}, which is normalized as
$\langle {f}(r,t) {f}(r',t') \rangle \approx (4\pi  16 \xi^2\lambda_L^2 \tau_{\psi}T_c/\Phi^2_0) \delta(r-r')\delta(t-t')$, where $\lambda_L$ is the London penetration
depth and $\Phi_0$ is a magnetic flux quantum.
The origin of the coordinate system is chosen in the center of the sample, so that $\{x,y\}\in [-L/2, L/2]$.

\begin{figure*}[] 
\centering
\begin{minipage}[h]{1.0\linewidth}
\includegraphics[width=0.6\textwidth]{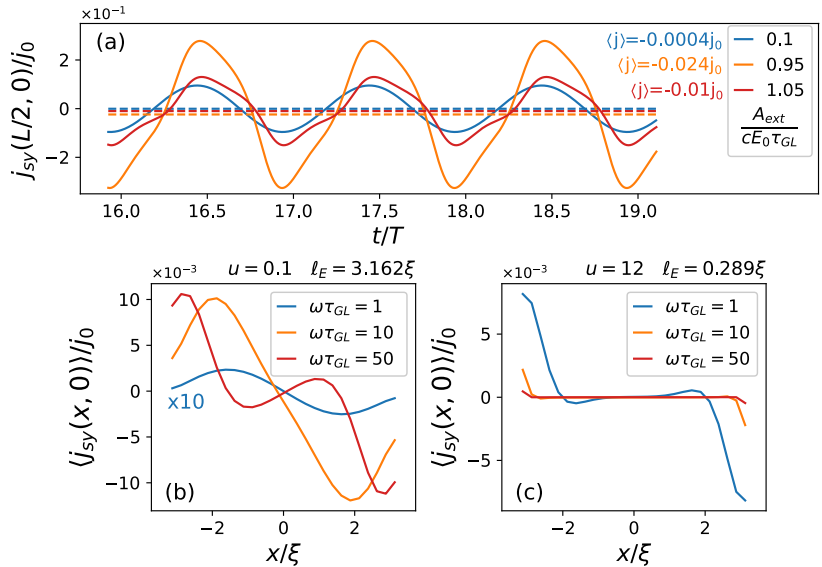} 
\end{minipage}
\caption{ {\small 
(a) Steady-state oscillations of the supercurrent $j_{sy}(x=L/2, y=0,t)$ in the square superconductor with a side $L=7\xi$ with a nonzero average $\langle j_{sy}\rangle_T$ (dashed lines) under the radiation of an external field with a frequency $\omega\tau_{\text{GL}}=2$ (period $T=\pi\tau_{\text{GL}}$) for parameters $u=1$, $\eta=0.3$.
(b, c) Spatial distribution of the averaged supercurrent $\langle j_{sy}(x,y=0)\rangle_T$ along a central section of the superconductor for different values of the electric field penetration length $\ell_E$ and frequency $\omega$ for  $\eta=0.3$ ; the amplitude of the external field is different for the different frequencies with the fixed relation $E_{\text{ext}}/\omega=A_{\text{ext}}/c=0.75(E_0\tau_{\text{GL}})$. 
}}
\label{fig1}
\end{figure*}

For a rather small sample of the size $L$ much less than the wavelength of the electromagnetic radiation one can assume the electric field ${\bf E}_{\text{ext}}({\bf r},t)$ of the electromagnetic wave to be uniform over the superconductor (see Fig. \ref{fig3}(a)), and the corresponding time-dependent dimensionless vector potential can be written in the form:   
$$
{\bf A}_{\text{ext}}(t)=\text{Re}\big[-icE_{\text{ext}}/\omega ({\bf x}_0 + \sigma_{\pm} i {\bf y}_0)e^{-i\omega t} \big].
$$
Here $c$ is the speed of light and the circular polarization is defined as $\sigma_{\pm}=\pm 1$ for different helicity of the electromagnetic wave. 
Dimensional unit for an electric field amplitude is $E_0=\hbar/2e\tau_{\text{GL}}\xi$, for a supercurrent and magnetic moment per unit area is $j_0=M_0c=\sigma_nE_0$, where $\sigma_n$ is a conductivity of a superconductor in a normal state and $c$ is the speed of light. 
We consider the case of small lateral sizes $L\ll\lambda_L^2/d$, where $d$ is the sample thickness, therefore we can neglect the contribution to the magnetic field induced by the supercurrents. 
This condition allows us to treat the function ${\bf A}_{\text{ext}}(t)$ as an external source in the Eqs. (\ref{GL1}, \ref{GL2}) by using a direct substitution ${\bf A}\equiv {\bf A}_{\text{ext}}$.

Numerical calculation is implemented as follows:
for each moment of time, the Poisson's equation (\ref{GL2}) for the potential $\varphi$ is solved using the Fourier method; then using the value of $\varphi(x,y,t)$ we find the order parameter from Eq. (\ref{GL1}) in the next time step $\psi(x,y,t+\Delta t)$ using the semi-implicit Crank-Nicolson scheme.

\section{Stationary regime of IFE}
First, we address the stationary case - without heating and quench dynamics.
An alternating harmonic electric field of a circularly polarized THz radiation of frequency $\omega$ induces a supercurrent with the density ${\bf j}_s({\bf r}, t)=\text{Re}\sum_{n} {\bf j}_{s,n}({\bf r})e^{in\omega t}$ including all harmonics $n\omega$ with an integer $n$. 
Note here that the even-$n$ harmonics in the superconducting condensate response appear only for a nonzero imaginary part of the order parameter relaxation time: $\eta\neq 0$.
An example of the multi-harmonic oscillations of the supercurrent ${\bf j}_s$ is shown in Fig. \ref{fig1}(a). 
According to the IFE theory for the superconducting condensate \cite{PhysRevLett.126.137002}, the same parameter $\eta$ is responsible for a nonzero averaged supercurrent induced by the electromagnetic wave:
\begin{gather}
\langle {\bf j}_s({\bf r})\rangle_T=\frac{1}{T}\int_0^{T}{\bf j}_s({\bf r},t)dt,
\end{gather}
where $T=2\pi/\omega$ is a period of the electric field. 
The direction of the current flow is determined by the helicity of the circular polarized wave $\sigma_{\pm}$.
The spatial distribution of the dc current is controlled by two characteristic length scales: (i) the electric field penetration length \cite{kopnin2009theory, IvlevKopnin} 
$\ell_E=\xi/\sqrt{u}$; (ii) the phenomenological frequency-dependent length $\ell_{\omega}\sim \xi/\sqrt{\omega}$.  
While the first length $\ell_E$ is the length of conversion of normal currents to the superconducting ones, the $\ell_{\omega}$ value can be qualitatively considered as a localization scale of the order parameter amplitude and phase \cite{PhysRevLett.126.137002}.
It is worth noting that the applicability of the TDGL model for externally driven processes is provided by a condition $\omega\tau_{\text{GL}}<T_c/(T_c-T_0)$, where $T_c$ is a critical temperature of a superconductor and $T_0$ is a substrate temperature. From the general constraint on the time variation of the order parameter $\omega_{\text GL}\equiv \tau_{\text {GL}}^{-1}\ll\tau_{\varepsilon}^{-1}$, where $\tau_{\varepsilon}$ is the inelastic relaxation time of quasiparticles \cite{kopnin2009theory, IvlevKopnin}, we obtain an additional condition $\omega \ll T_c/(T_c-T_0)\tau_{\varepsilon}^{-1}$, which is always satisfied in the vicinity of $T_c$.

The distribution of $y$-component of the dc current $\langle {\bf j}_s\rangle_T$ induced by the radiation with $\sigma_{+}$ polarization for different values of $\omega$ and $u$ is presented in Fig. \ref{fig1}. 
It is straightforward from the fourfold symmetry of the problem that the distribution $\langle j_{sx}(x=0,y)\rangle_T$ can be obtained by $\pi/2$ rotation.
For convenience, the amplitude of the time-dependent vector potential $A_{\text{ext}}=cE_{\text{ext}}/\omega$ is fixed for all plots: $A_{\text{ext}}=0.75(cE_0\tau_{\text{GL}})$.
We observe that the transition from the adiabatic $\omega\tau_{\text{GL}}\lesssim 1$ to the nonadiabatic $\omega\tau_{\text{GL}}\gg 1$ regime is accompanied by the strong decrease of the localization length of the supercurrent $\langle {\bf j}_s({\bf r})\rangle_T$  for $u\ll 1$ and rather weak decrease for $u\gg 1$.
The localization length of the supercurrent is determined by the frequency $\omega$ when the largest length scale in the superconductor is $\ell_E\sim L$ and mainly by the parameter $u$ when this length scale is $\ell_{\omega} \lesssim L$ (see Fig. \ref{fig1}). 
Therefore the supercurrent is always localized at the smallest length scale $\sim\text{min}\{\ell_E, \ell_{\omega}\}$.

\section{Magnetic moment}
\begin{figure*}[] 
\centering
\begin{minipage}[h]{1.0\linewidth}
\includegraphics[width=0.65\textwidth]{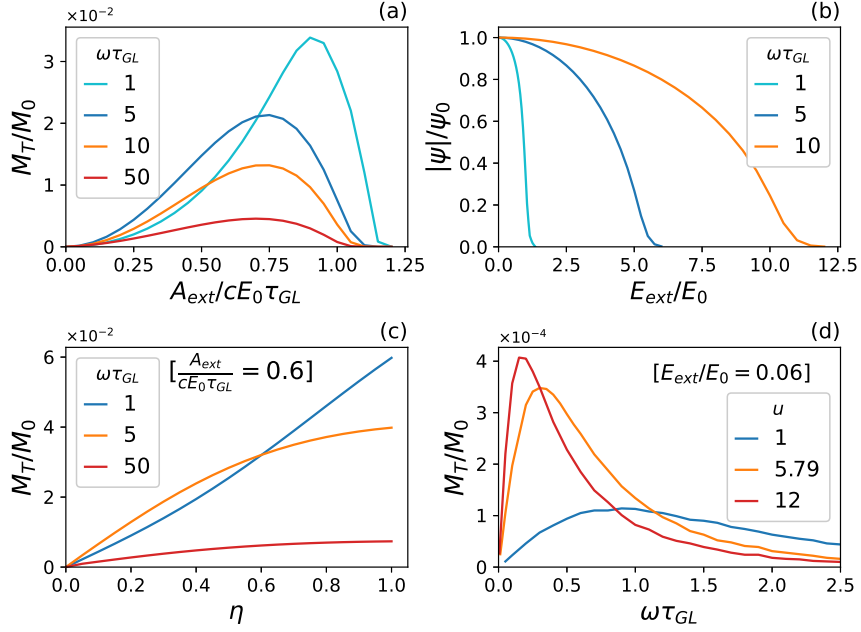} 
\end{minipage}
\caption{ {\small Dependence of the dc magnetic moment $M_T$ in the square superconductor with a side $L=7\xi$ on different parameters (a, c, d). 
Subplot (b) shows the dependence of the amplitude of the order parameter $|\psi|(x=0,y=0)$ on the $E_{\text{ext}}$ and corresponds to (a). 
Sets of parameters are chosen as follows:
(a, b) $u=1$, $\eta=0.3$;
(c) $u=1$, $A_{\text{ext}}=0.6(cE_0\tau_{\text{GL}})$; 
(d) $\eta=0.3$, $E_{\text{ext}}=0.06E_0$; 
 }}
\label{fig2}
\end{figure*}

The averaged current $\langle {\bf j}_s\rangle_T$ produces a dc magnetic moment per unit area
$ {\bf M}_T=L^{-2}\int[{\bf r}\times \langle{\bf j}_s({\bf r})\rangle_T]d{\bf r}$, with a direction determined by the light polarization. 
The dependencies of the modulus of the magnetic moment $|{\bf M}_T|\equiv M_T$ on different parameters are shown in Fig. \ref{fig2}.
Figure \ref{fig2}a demonstrate that with a small amplitude of the vector potential, the moment grows quadratically as $M_T\sim A_{\text{ext}}^2$ and after passing the maximum value at $A_{\text{ext}}\approx 0.75(cE_0\tau_{\text{GL}})$ the moment begins to decrease due to the suppression of the order parameter $\psi$, shown in Fig. \ref{fig2}(b). 
The term $|{\bf A}|^2$ in the TDGL equation for the order parameter can be treated as a negative contribution to the critical temperature $T_c$, therefore superconductivity is destroyed and the moment decreases to zero at $A_{\text{ext}}\approx 1.0(cE_0\tau_{\text{GL}})$, which corresponds to $E_{\text{ext}}\approx\omega\sqrt{0.5H_{c2}(T_0)\Phi_0}/c^2$ in dimensional units.
At low frequencies and large amplitudes of the external field, the distribution of the order parameter and, correspondingly, the supercurrent ${\bf j}_s$ become strongly inhomogeneous,  which leads to a shift of the maximum of $M_T(E_{\text{ext}})$ at $\omega\tau_{\text{GL}}=1$  relative to the curves plotted for larger frequencies.
The moment $M_T$ as a function the frequency $\omega$ has a peak shown in Fig. \ref{fig2}(d).
For the fixed amplitude of the external filed $E_{\text{ext}}$ the moment grows linearly $M_T\sim \omega$ at $\omega\tau_{\text{GL}}\ll 1$ and decreases as $M_T\sim \omega^{-3}$ at $\omega\tau_{\text{GL}}\gtrsim 1$. 
This behavior is in a good agreement with the perturbative analytical solution provided in \cite{PhysRevLett.126.137002}. 
Using the optimal parameters one can achieve the most efficient interaction of the dc current produced by IFE with Abrikosov vortices generated by the thermal quench, which is discussed below. 

\section{Numerical simulation of the vortex generation}
\begin{figure*}[] 
\begin{minipage}[h]{1.0\linewidth}
\includegraphics[width=1.0\textwidth]{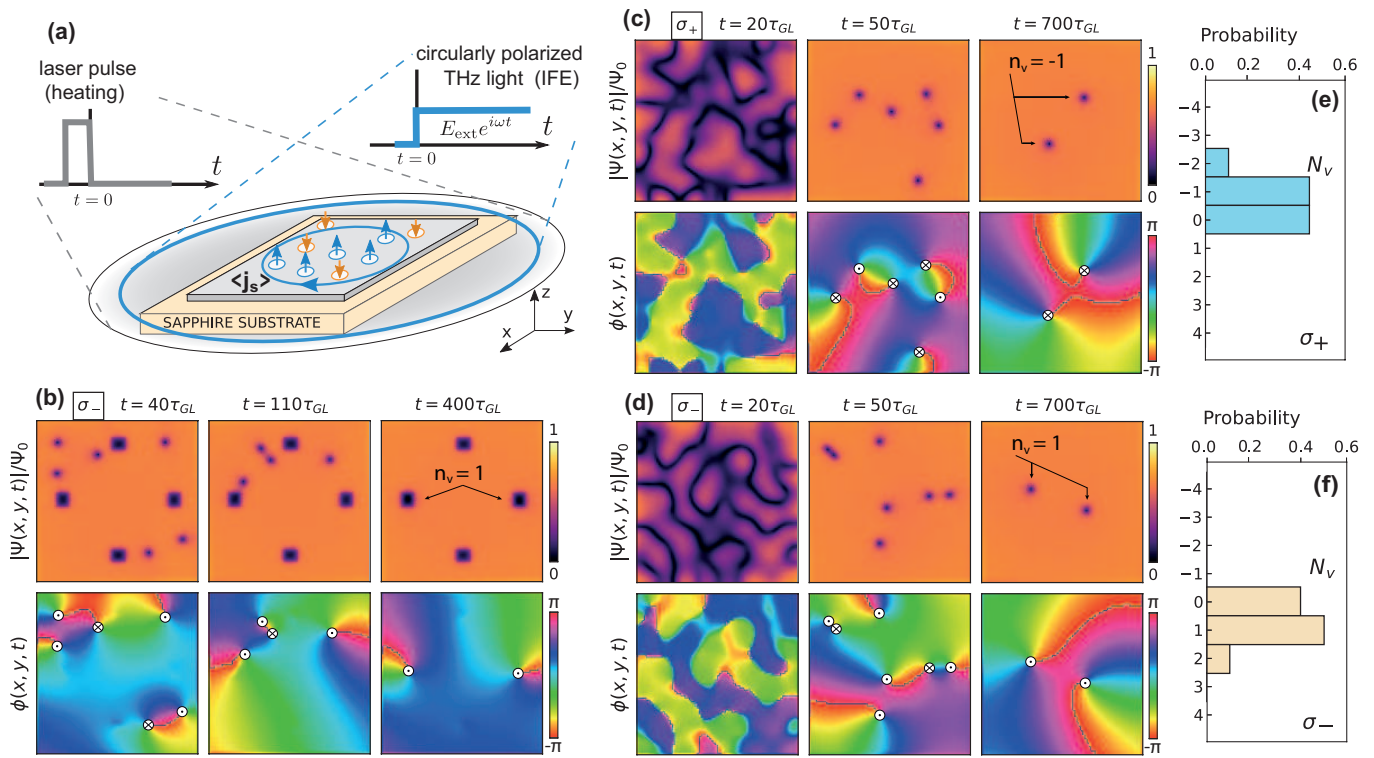}
\end{minipage}
\caption{ {\small (a) Sketch of the proposed experimental setup: superconductor placed on a sapphire substrate is heated by an external laser pulse and then quenched in the presence of a circularly polarized light. (Orange)blue arrows show (anti-)vortices created after the rapid thermal quench.
(c, d) Numerical simulations of a vortex nucleation and dynamics in the presence of a circularly polarized light with $\sigma_+$ and $\sigma_-$ polarization. 
(b) Pinning of the vortex by a small square defect after vortex nucleation. 
Panels (b-d) show the modulus $|\psi|$ and the phase $\phi$ of the order parameter for different time instants. White circles with a dot(cross) denote vortices with a polarity $n_{\text{v}}=1 (-1)$. 
(e, f) Probability of the creation of the vortices with a certain vorticity the for different polarizations $\sigma_+$ and $\sigma_-$. 
The number of the implementations (unique simulations) is $N_{\text{imp}}=20$ for each subplot.
The set of parameters used for the calculation (b-f): $L=80\xi$, $u=1$,  $\omega\tau_{\text{GL}}=10$, $\eta=0.3$ , $E_{\text{ext}}=7.5E_0$, $\tau_q=1.0\tau_{\text{GL}}$.}}
\label{fig3}
\end{figure*}

In order to implement the optical generation of the Abrikosov vortices with a desired polarity, we consider a process consisting of two subsequent illumination stages (see Fig.\ref{fig3}(a)): (i) before the time instant $t=0$ the superconductivity  in the film is completely destroyed due to the sample heating by a strong laser pulse with the beam radius well exceeding the size $L$; (ii) a rapid thermal quench occurs
at the second stage for $t>0$ in the presence of a weak circularly polarized electromagnetic wave.
We assume that the temperature distribution over the film is uniform and its time evolution can be described by the phenomenological expression \cite{PhysRevB.63.184520}:  $T(t)=T_0+(T_i-T_0)e^{-t/\tau_q}$, where $T_i$ is an initial temperature of the superconductor and $\tau_q$ is a characteristic heat drain time.
Following this model and taking $T_i>T_c$ we assume the homogeneous initial conditions $\psi({\bf r}, t=0)=0$.
After the start of the quench at $ t = 0 $, superconductivity begins to recover in the presence of the thermal fluctuations $f({\bf r},t)$ and, according to the Kibble-Zurek mechanism, the vortex-antivortex pairs appear throughout the sample. 
Further dynamics of these vortex pairs at times $t\gg\tau_q$ is affected by the circularly polarized radiation with the frequency $\omega$.
Since the induced current has both dc and ac components, the equation of the motion for the vortex has a quite complicated form. 
The alternating electric field produces local oscillations of the vortex position which are observable at $\omega\lesssim \tau^{-1}_{\text{GL}}$ and are averaged at larger frequencies.
The averaged part of the current produces a Lorentz force $f_L\sim\langle j \rangle_T$ acting on a single vortex. 
The direction of the force is defined both by the sign of the polarization $\sigma_{\pm}$ and the vortex winding number, or polarity:
$
n_{\text{v}}=\frac{1}{2\pi}\oint_l\nabla\arg(\psi)dl,
$
where $l$ is the anticlockwise oriented contour around a single vortex. In the following we use the term 'vortex' for $n_{\text{v}}=1$ and 'antivortex' for $n_{\text{v}}=-1$.
In the presence of the imaginary part of the relaxation time $\eta\neq 0$ there is a Hall component of the vortex motion (see, e.g., \cite{kopnin2009theory}).
Phenomenologically one can describe this effect as 
$$
\sigma n_{\text{v}}\langle {\bf j}_s\rangle_T\times{\bf z}_0=\alpha_1{\bf v}_L+n_{\text{v}}\alpha_2(\eta){\bf v}_L\times {\bf z}_0,
$$
where ${\bf v}_L$ is the local vortex velocity, $\alpha_1{\bf v}_L$ is the viscous drag force and $\alpha_2$ corresponds to the Hall effect. It is useful to note that $\alpha_2(\eta=0)=0$ \cite{PhysRevB.46.8376, kopnin2009theory,Kopnin_sign}. 

The dynamics of the superconducting condensate during the quench process are presented in Fig. \ref{fig3}(c,d) (see Supplementary movies \cite{SM_movies}).
At the initial stage $t\approx 20\tau_{\text{GL}}$ we observe the nucleation of the vortex-antivortex pairs which are distributed randomly over the superconductor area since the quench is homogeneous.
After that at $t\approx 50\tau_{\text{GL}}$ the part of the pairs annihilates and the remaining (anti)vortices begin to move in the presence of the induced supercurrent. 
Since the (anti)vortex-current interaction depends on the direction of the dc supercurrent, the current acts selectively expelling the vortices with a certain polarity from the sample. 
It is shown in Fig. \ref{fig3}(c,d) that for the $\sigma_{\pm}$ polarization only antivortices/vortices with $n_{\text{v}}=\mp1$ survive in the sample at the times $t\sim 700\tau_{\text{GL}}$. 
In the absence of the pinning surviving (anti)vortices stay in the superconductor for quite a long time: they escape from the superconductor only for $t\gg 700\tau_{\text{GL}}$. 
The formation of vortex-antivortex pairs is controlled by a stochastic force and in order to establish the correlation between a given polarization and the polarity of vortices surviving at large times one needs to consider a statistical dependence $N_{\text{v}}(\sigma)$, where $N_{\text{v}}=\sum n_{\text{v}}$ is the sum over all vortices.
This dependence is shown in Fig. \ref{fig3}(e,f) and it is clearly seen that $N_v(\sigma_+)\leq0$ and $N_v(\sigma_-)\geq0$ for $N_{\text{imp}}=20$ implementations for each polarization. 
Obviously, the distributions for $\sigma_{+}$ and $\sigma_{-}$ should be symmetric in the limit $N_{\text{imp}}\rightarrow\infty$.
Note that among the results of calculations we also observe the implementations with $N_v=0$ when all the vortices and antivortices are either annihilated or leaving the sample for the times $t\sim 700\tau_{\text{GL}}$. 
In the case of linear polarization, which is the sum of two waves with opposite helicities, we observe only $N_v=0$ for the times $t\gtrsim 700\tau_{\text{GL}}$, since the IFE is absent.

Obviously, the escape of vortices from the superconductor can be additionally prevented by introduction of pinning centers. 
In order to strengthen the influence of pinning we should place these centers near the edges of the superconductor, where the dc supercurrent $\langle {\bf j}_s\rangle_T$ is maximal and plays, thus, a stronger role in separation of vortex-antivortex pairs (see Fig. \ref{fig2}(b,c)).
An example of such a process is shown in Fig. \ref{fig3}(b) for the case of the square defects with locally suppressed superconducting critical temperature $T_c$ \cite{SM_movies}. 
Numerical simulation shows that the polarity of the pinned vortices is consistent with the helicity of the light polarization, according to the statistical dependence $N_{\text{v}}(\sigma)$. 
These observations prove the possibility of creation of vortices with a certain polarity in the absence of the applied magnetic field only by the circularly polarized electromagnetic wave.
Generated vortices contribute to the dc magnetic moment providing, thus, a possibility to observe the enhanced IFE.

\section{Discussion}
Reduction in the parameter $\eta$, used in the simulation of the vortex dynamics above, leads to a decrease in the amplitude of the averaged current (see Fig. \ref{fig2}(c)), which makes locking of a vortex with a desired polarity less likely. 
Therefore, an experimental observation of the light-induced vortex generation is possible in materials with relatively large imaginary part of the superconducting relaxation time $\eta\sim T_c/E_F\lesssim 1$. Since the parameter $\eta$ is also responsible for the Hall effect and the Hall-anomaly in the vortex state of type-II superconductors \cite{Kopnin_sign, PhysRevB.46.8376,kopnin2009theory}, promising candidates for an experiment can be high-$T_c$ compounds, where studies indicate relatively large Hall effect \cite{PhysRevLett.122.247001, PhysRevB.43.6246, PhysRevB.104.L020503, PhysRevB.49.4209, TINH201629}. 
Among other possible candidates with quite a large relation $T_c/E_F\sim 0.3$ one can mention the class of actively studied iron selenides \cite{sym12091402, doi:10.7566/JPSJ.89.102002}.

Consider a specific example of a thin YBCO sample with the size $L\sim 0.1-4$ $\mu$m and $d\sim 10$nm. 
For the substrate temperature $T_0\approx0.98T_c$ (with $T_c\approx 90$ K) the typical frequency of the circularly polarized radiation used in the calculation corresponds to the far infrared range $\omega\sim 10/\tau_{\text{GL}} \sim 50$ THz. 
Corresponding intensity of the polarized radiation at which the effect is the most pronounced is $\mathcal{I} \approx 5\cdot 10^{-2} $ $\mu$W/$\mu$m$^2$. 
Note, that low temperature materials Nb or FeSe with $T_c\sim 9$ K require a terahertz frequency range $\omega\sim 1-10$ THz.
The control of the quench time $\tau_q$ in an experiment is possible due to good heat removal from the superconductor ensured by, for example, sapphire substrate film (see Fig. \ref{fig1}) with a typical thickness $\sim 1$ $\mu$m \cite{doi:10.1021/acs.nanolett.0c02166}. It provides large thermal conductivity $\sim 10^3 $ W/mK \cite{berman1955thermal}, which ensures the heat transfer of a surface power density of the order of $\sim 10$ $\mu$W/$\mu$m$^2$ at the temperature change of the superconductor $\Delta T\sim10^{-2}$ K.

Vortex polarity can be detected with the local vortex imaging provided by the SQUID measurements with sub-micron spatial resolution \cite{doi:10.1063/1.4961982,JoseMartinezPerezKoelle2017, Vasyukov2013}, the scanning magnetometry with nitrogen-vacancy centers in diamonds \cite{Thiel2016,Lenz_2021}, the magneto-optical imaging based on Faraday rotation of light polarization \cite{Goa_2001, doi:10.1021/acs.nanolett.0c02166, Veshchunov2016} or magnetic force microscopy technique \cite{Suderow_1,PhysRevResearch.2.013329}. 
It is also possible to use an array of superconducting disks \cite{doi:10.1126/science.1178139} simultaneously irradiated with polarized radiation, while the average magnetic moment can be measured using a standard SQUID magnetometer technique \cite{Kirtley_2010}.

In summary, we theoretically showed that vortex-antivortex pairs created by a thermal laser pulse in a superconductor can be separated by the dc supercurrent induced by an external circularly polarized radiation due to IFE. This leads to effective locking of vortices with a certain polarity inside the superconductor, determined by the light polarization.
The findings of this research can be applied in experiments on a fast vortex manipulation in mesoscopic superconductors.

\section{Acknowledgements} 
The work has been supported by the ANR OPTOFLUXONICS, the LIGHT S$\&$T Graduate Program, the Russian Science Foundation (Grant No. 21-72-10161) and the IdEx of the University of Bordeaux/Grand Research Program "GPR LIGHT".


%


\end{document}